\documentclass[12pt]{article}

\usepackage{a4wide}
\usepackage[pdftex,usenames,dvipsnames]{color}
\usepackage{graphics}
\usepackage{amsfonts}

\newcommand{\nit}{\noindent}

\newcommand{\np}{\newpage}
\newcommand{\dsp}{\displaystyle}
\newcommand{\vs}[1]{\vspace{#1 ex}}
\newcommand{\hs}[1]{\hspace{#1 em}}
\newcommand{\bfr}{\begin{flushright}}
\newcommand{\efr}{\end{flushright}}
\newcommand{\bc}{\begin{center}}
\newcommand{\ec}{\end{center}}
\newcommand{\ben}{\begin{enumerate}}
\newcommand{\een}{\end{enumerate}}

\newcommand{\be}{\begin{equation}}
\newcommand{\ee}{\end{equation}}
\newcommand{\ba}{\begin{array}}
\newcommand{\ea}{\end{array}}
\newcommand{\ct}{\cite}
\newcommand{\bit}{\bibitem}
\newcommand{\dd}[2]{\frac{\partial{#1}}{\partial{#2}}}
\newcommand{\ag}{\alpha}

\newcommand{\gam}{\gamma}
\newcommand{\del}{\delta}

\newcommand{\ve}{\varepsilon}

\newcommand{\kg}{\kappa}
\newcommand{\lb}{\lambda}
\newcommand{\sg}{\sigma}

\newcommand{\og}{\omega}
\newcommand{\Gam}{\Gamma}
\newcommand{\Del}{\Delta}
\newcommand{\Fg}{\Phi}

\newcommand{\bfv}{\bold{v}}

\newcommand{\bfB}{\bold {B}}
\newcommand{\bfE}{\bold {E}}

\newcommand{\lh}{\left(}
\newcommand{\rh}{\right)}
\newcommand{\ld}{\left.}
\newcommand{\rd}{\right.}

\newcommand{\der}{\partial}

\begin{document}

\pagestyle{empty}

\bfr
\efr
\vs{5}

\bc
{\bf \large  The gravity of light-waves}$^*$ 
\vs{4} 

{\large J.W.\ van Holten}
\vs{3} 

Nikhef, Amsterdam \\
    and \\
Leiden University \\
Netherlands
\ec

\nit
{\small 
{\bf Abstract} \\
Light waves carry along their own gravitational field; for simple plain electromagnetic waves the 
gravitational field takes the form of a $pp$-wave. I present the corresponding exact solution 
of the Einstein-Maxwell equations and discuss the dynamics of classical particles and 
quantum fields in this gravitational background.}

\vfill
\nit
\footnoterule
\nit
$^*$ {\footnotesize Lecture presented at the meeting {\em Estate Quantistica} 2018, Scalea (Italy) } \\

\np
\pagestyle{plain} 
\pagenumbering{arabic} 

\nit
{\bf \large 1.\ Setting the stage} 
\vs{1}

\nit
The gravitational properties of light waves have been studied extensively in the literature 
\ct{tolman:1931}-\ct{lynden-bell:2017}. In this lecture I describe the exact solutions 
of Einstein-Maxwell equations discussed in \ct{jwvh:1999,jwvh:2008} and some applications.

The discussion concerns plain electromagnetic waves propagating in a fixed direction chosen to be 
the $z$-axis of the co-ordinate system. As they propagate at the universal speed $c$, taken to be
unity: $c =1$ in natural units, it is useful to introduce light-cone co-ordinates $u = t - z$, $v = t + z$.
Then the electromagnetic waves to be discussed are described by a transverse vector potential
\be
A_i(u) = \int \frac{dk}{2\pi}\, \lh a_i(k) \sin ku + b_i(k) \cos ku \rh, \hs{2} i = (x, y).
\label{1.1}
\ee
This expression explicitly makes use of the superposition principle for electromagnetic fields, 
guaranteed in Minkowski space by the linearity of Maxwell's equations and well-established 
experimentally. The corresponding minkowskian energy-momentum tensor is 
\be
T_{\mu\nu} = F_{\mu\lb} F_{\nu}^{\;\,\lb} - \frac{1}{4} \eta_{\mu\nu} F_{\kg\lb} F^{\kg\lb},
\label{1.2}
\ee
the only non-vanishing component of which in light-cone co-ordinates is 
\be
T_{uu} = \frac{1}{2} \lh \bfE^2 + \bfB^2 \rh.
\label{1.3}
\ee
Here the components of the transverse electric and magnetic fields are expressed in terms of 
the vector potential (\ref{1.1}) by
\be
E_i(u) = - \ve_{ij} B_j(u) = A'_i(u). 
\label{1.4}
\ee
the prime denoting a derivative w.r.t.\ $u$. 

The same expression for light-waves also holds in general relativity, the corresponding 
special solution of the Einstein equations being described by the line element 
\be
ds^2 = - du dv - \Fg(u, x, y) du^2 + dx^2 + dy^2
\label{1.5}
\ee
For this class of metrics \ct{brinkmann:1923,baldwin-jeffery:1926} the only non-vanishing 
components of the connection are 
\be
\Gam_{uu}^{\;\;\;v} = \der_u \Fg, \hs{2} \Gam_{iu}^{\;\;\;v} = 2 \Gam_{uu}^{\;\;\;i} = \der_i \Fg,
\label{1.6}
\ee
and the complete Riemann tensor is given by the components 
\be
R_{uiuj} = - \frac{1}{2}\, \der_i \der_j \Fg.
\label{1.7}
\ee
As a result the Ricci tensor is fully specified by 
\be
R_{uu} = - \frac{1}{2} \lh \der_x^2 + \der_y^2 \rh \Fg,
\label{1.8}
\ee
which matches the form of the energy-momentum tensor (\ref{1.3}) and thus allows 
solutions of the Einstein equations specified by 
\be
\Fg = 2 \pi  G (x^2 + y^2) \lh \bfE^2 + \bfB^2 \rh(u) + \Fg_0(u,x^i),
\label{1.9}
\ee
with $\Fg_0$ representing a free gravitational wave of $pp$-type. 
\np
\nit
{\bf \large 2.\ Geodesics}
\vs{1}

\nit
The motion of electrically neutral test particles in a light-wave (\ref{1.1}) is described by the 
geodesics $X^{\mu}(\tau)$ of the $pp$-wave space-time (\ref{1.5}). They are found by 
solving the geodesic equation
\be
\ddot{X}^{\mu} + \Gam_{\lb \nu}^{\;\;\;\mu} \dot{X}^{\lb} \dot{X}^{\nu} = 0,
\label{2.1}
\ee
the overdot denoting a derivative w.r.t.\ proper time $\tau$. The equation for the geodesic 
light-cone co-ordinate $U(\tau)$ is especially simple, as its momentum (representing a 
Killing vector) is conserved:
\be
\dot{U} = \gam = \mbox{constant}.
\label{2.2}
\ee
Another conservation law is found from the hamiltonian constraint obtained by substitution of the
proper time in the line element:
\be
-1 = - \dot{U} \dot{V} - \Fg(U,X^i) \dot{U}^2 + \dot{X}^{i\,2} 
\hs{1} \Leftrightarrow \hs{1} \frac{1}{\gam^2} = \frac{1-\bfv^2}{(1 - v_z)^2} + \Fg,
\label{2.3}
\ee
where $\bfv = d{\bf X}/dT$ is the velocity in the observer frame.
Finally, using (\ref{2.2}) to substitute $U$ for $\tau$, the equations for the transverse
co-ordinates become 
\be
\frac{d^2 X^i}{dU^2} + \frac{1}{2}\, \dd{\Fg}{X^i} = 0.
\label{2.4}
\ee
For quadratic $pp$-waves $\Fg(u,x^i) = \kg_{ij}(u)\, x^i x^j$ this takes the form of 
a parametric oscillator equation
\be
\frac{d^2 X^i}{dU^2} + \kg_{ij}(U) X^j = 0.
\label{2.5}
\ee
For light-like geodesics the equations are essentially the same, except that the 
hamiltonian constraint is replaced by 
\be
\frac{1-\bfv^2}{(1 - v_z)^2} + \Fg = 0.
\label{2.6}
\ee
Note that in Minkowski space, where $\Fg = 0$, this reduces to $\bfv^2 = c^2 = 1$.
These equations take a specially simple form for circularly polarized light waves 
sharply peaked around a central frequency
\be
A_x(u) = \int \frac{dk}{2\pi}\, a(k) \cos ku, \hs{2} A_y(u) = \int \frac{dk}{2\pi}\, a(k) \sin ku,
\label{2.7}
\ee
where the domain of $a(k)$ is centered around the value $k_0$ with width $\Del k$ and 
central amplitude $a_0$. Then 
\be 
\mu^2 \equiv 2 \pi G \lh \bfE^2 + \bfB^2 \rh = 2 G \int dk\, k^2 a^2(k) 
 \sim G \Del k\, k_0^2\, a_0^2.
\label{2.8}
\ee
and therefore
\be
\Fg = \mu^2 \lh x^2 + y^2 \rh, \hs{2} \mu^2 = 4 \pi G\, \int \frac{dk}{2\pi}\, k^2 a^2(k).
\label{2.9}
\ee
Then equation (\ref{2.5}) reduces to a simple harmonic oscillator equation with angular 
frequency $\mu$ in the $U$-domain.

\np
\nit
{\bf \large 3.\ Field theory}
\vs{1}

\nit
In the previous sector we studied the equation of motion of test particles, supposed to 
have negligible back reaction on the gravitational field described by the metric (\ref{1.5}). 
Similarly one can study the dynamics of fields in this background space-time in the 
limit in which the fields are weak enough that their gravitational back reaction can 
be neglected. First we consider a scalar field $\Psi(x)$ described by the Klein-Gordon 
equation
\be
\lh - \Box_{pp} + m^2 \rh \Psi = 0, \hs{2} 
\Box_{pp} = - 4 \der_u \der_v + 4 \Fg(u,x^i) \der_v^2 + \der_x^2 + \der_y^2.
\label{3.1}
\ee
It is convenient to consider the Fourier expansion w.r.t.\ the light-cone variables $(u,v)$:
\be
\Psi(u,v,x^i) = \frac{1}{2\pi}\, \int ds dq\, \psi(s,q,x^i) e^{-i(su + qv )}.
\label{3.2}
\ee
Note that 
\be
su + qv = Et - pz, \hs{2} E = s + q, \hs{1} p = s - q.
\label{3.3}
\ee
Then the amplitudes $\psi$ satisfy the equation 
\be
\left[ \der_x^2 + \der_y^2 + 4 sq - 4 q^2 \Fg(- i\der_s, x^i) - m^2 \right] \psi = 0.
\label{3.4}
\ee
This equation can be solved explicity for the circularly polarized wave packets
which lead to the simple quadratic amplitude (\ref{2.9}). Then
\be
 \lh 4 sq - m^2 \rh \psi = \lh - \der_x^2 - \der_y^2 + 4 \mu^2 q^2 (x^2 + y^2) \rh \psi.
\label{3.5}
\ee
The right-hand side describes a couple of quantum oscillators with frequency $\og = 2 \mu |q|$
possessing an eigenvalue spectrum 
\be
2 \mu |q| \lh n_x + n_y + 1 \rh \equiv 4 \sg |q| , \hs{2} n_i = 0, 1, 2, ...
\label{3.6}
\ee
Thus equation (\ref{3.5}) reduces to
\be
\ba{l}
4 s q - 4 \sg |q| = m^2 \hs{.5} \mbox{or} \hs{.5}
\left\{ \ba{ll} (E - \sg)^2 = (p -\sg)^2 + m^2, & q > 0; \\
                   (E + \sg)^2 = (p+\sg)^2 + m^2, & q < 0. \ea \rd
\ea
\label{3.7}
\ee
The final result for the scalar field then becomes
\be
\ba{l}
\dsp{ \Psi(u,v,x^i) = \frac{1}{2\pi}\,  \int_0^{\infty} \frac{dq}{\sqrt{q}}\, \sum_{n_i = 0}^{\infty} 
 \lh a_{n_i}(q) e^{-iqv - i \lh \frac{m^2}{4q} + \sg \rh u} + a^*_{n_i}(q) e^{iqv +  i \lh \frac{m^2}{4q} + \sg \rh u} \rh }\\
 \\
\dsp{ \hs{7} \times\, \sqrt{ \frac{2\mu}{\pi} }\, \prod_{j=x,y} 
     \left[ \frac{H_{n_j}(\xi_j)}{\sqrt{2^{n_j} n_j!}}\, e^{- \xi_j^2} \right], \hs{2} \xi_j = \sqrt{2\mu q}\, x_j. }
\ea
\label{3.8}
\ee

\np
\nit
{\bf \large 4.\ Electromagnetic fluctuations in a light-wave background} \\
\vs{1}

\nit
On top of an electromagnetic wave described by equation (\ref{1.1}) there can be fluctuations 
of the electromagnetic field. The general form of the Maxwell field then is of the form
\be
A_{\mu}(u,v,x^i) = \del_{\mu}^i A^{wave}_i(u) + a_{\mu}(u,v,x^i).
\label{4.1}
\ee
Because of the linearity of Maxwell's equations the field equations for the wave background 
and the fluctuations separate. The fluctuating field equations in the gravitational $pp$-wave 
background are derived from the action 
\be
\ba{lll}
S & = & \dsp{ \int dudvdxdy \left[  \frac{}{} \lh \der_u a_v - \der_v a_u \rh^2 + \lh \der_u a_i - \der_i a_u \rh 
 \lh \der_v a_i - \der_i a_v \rh - \Fg \lh \der_v a_i - \der_i a_v \rh^2 \rd }\\
 & & \\
 & & \dsp{ \hs{6} \ld -\, \frac{1}{8}  \lh \der_i a_j - \der_j a_i \rh^2 \right], }
\ea
\label{4.2}
\ee
and read
\be
\ba{l}
\dsp{ \frac{\del S}{\del a_u} = 4 \der_v \der_u a_v - \Del_{\perp} a_v - 
  2 \der_v \lh \der_v a_u + \der_u a_v - \frac{1}{2}\, \der_i a_i \rh = 0, }\\
  \\
\dsp{ \frac{\del S}{\del a_v} = 4 \der_v \der_u a_u - \Del_{\perp} a_u -
 2 \der_u \lh \der_v a_u + \der_u a_v - \frac{1}{2}\, \der_i a_i \rh 
 + 2 \der_i \left[ \Fg \lh \der_i a_v - \der_v a_i \rh \right] = 0, }\\
 \\
\dsp{ \frac{\del S}{\del a_i} = - 2 \der_v \der_u a_i + \frac{1}{2}\, \Del_{\perp} a_i 
 + \der_i \lh \der_v a_u + \der_u a_v - \frac{1}{2}\, \der_j a_j \rh 
 - 2 \der_v \left[ \Fg \lh \der_i a_v - \der_v a_i \rh \right] = 0, }
\ea
\label{4.3}
\ee
where $\Del_{\perp} = \der_x^2 + \der_y^2$. As the fluctuating field equations possess their 
own gauge invariance they can be restricted without loss of generality by the constraint 
\be
 \der_v a_u + \der_u a_v - \frac{1}{2}\, \der_i a_i = 0.
\label{4.4}
\ee
However, this does not yet exhaust the freedom to make gauge transformations, as the 
condition (\ref{4.4}) is repected by special gauge transformations
\be
a'_{\mu} = a_{\mu} + \der_{\mu} \ag, \hs{1} \mbox{with} \hs{1}  \lh 4 \der_u \der_v - \Del_{\perp} \rh \ag = 0.
\label{4.5}
\ee
As can be seen from the first equation (\ref{4.3}) these transformations can be used to eliminate the 
component $a_v$ by taking 
\be
\der_v \ag = - a_v \hs{1} \Rightarrow \hs{1} a'_v = 0.
\label{4.6}
\ee
We are then left with a fluctuating field component $a_u$ restricted by (\ref{4.4}):
\be
\der_v a_u = \frac{1}{2}\, \der_i a_i,
\label{4.7}
\ee
implying $a_u$ to satisfy the Gauss law constraint 
\be
\der_i \left[ \der_i a_u - 2 \lh \der_u - \Fg \der_v \rh a_i \right] = 0.
\label{4.8}
\ee
The only remaining dynamical degrees of freedom are now the transverse components $a_i$ 
which are solutions of the Klein-Gordon type of equations
\be
\lh - 2 \der_u \der_v + 2 \Fg \der_v^2 + \frac{1}{2}\, \Del_{\perp} \rh a_i = 0.
\label{4.9}
\ee
For $pp$-backgrounds of the special form (\ref{2.9}) these solutions take the form (\ref{3.8}) 
with $m^2 = 0$. 

In the full theory also the gravitational field must fluctuate in a corresponding fashion. In the 
limit where the fluctuations are due to irreducible quantum noise, a corresponding quantum 
effect must be present in the space-time curvature. In view of the result (\ref{1.9}) for the 
photon fluctuations in the light-beam itself these are expected to take the form of associated 
spin-0 graviton excitations.

\end{document}